# MANAGEMENT OF A LARGE DISTRIBUTED CONTROL SYSTEM DEVELOPMENT PROJECT*

D.P. Gurd, SNS LANL/ORNL, Oak Ridge, TN 37830 USA
{*} Work supported by the US Department of Energy under contract DE-AC05-00OR22725

Abstract

Building an accelerator at six geographically dispersed sites is quite mad, but politically expedient. The Spallation Neutron Source (SNS), currently under construction in Oak Ridge, Tennessee, combines a pulsed 1 Gev $H^-$ superconducting linac with a compressor ring to deliver 2MW of beam power to a liquid mercury target for neutron production [1]. Accelerator components, target and experimental (neutron-scattering) instruments are being developed collaboratively by Lawrence Berkeley (Ion Source and Front End), Los Alamos (Linac), Thomas Jefferson (Cryosystems), Brookhaven (Compressor Ring), Oak Ridge (Target and Conventional Facilities) and Argonne (Neutron Scattering Instruments) National Laboratories. Similarly, a team distributed among all of the participating laboratories is developing the EPICS-based control system. This paper discusses the management model and strategies being used to address the unusual issues of organization, communication, standardization, integration and hand-off inherent in this widely-distributed project.

## 1 INTRODUCTION

The SNS control system presents no special or unique technical challenges. It is being developed using a standard, flat, EPICS-based architecture [2], using linux-based upper layer clients, consoles and servers; distributed VME- and VXI-based input-output controllers (IOCs) with Motorola 2100 series Power PC processors and a PLC or field-bus I/O layer for process control-like subsystems [3,4]. The timing and synchronization system is based upon RHIC hardware [5], which implements concepts originated at Fermilab perhaps twenty years ago. The PLC-based personnel safety system [6] is modeled upon a similar system in use at Jefferson Lab. The communication network is based upon now-standard switched Gigabit Ethernet [7]. The most original aspect of the SNS architecture is the use of PC-based "Network Attached Devices" (NADS), developed by the Beam Diagnostics team for beam instrumentation [8]. These NADS are designed to look to the control system like EPICS IOCs.

Implementing these more-or-less conventional systems using teams distributed across the country, belonging to laboratories each of which brings its own culture and approach, does however present a unique and interesting challenge. Conventional management approaches, organization, and communication methods must all be adapted to the realities of the partnership.

There have been successes and failures, but the effort is of interest because it seems likely that for political and economic reasons future large projects will also be built as collaborations. These projects may well be international in scope, and the issues thereby exacerbated. What is learned at SNS should be useful.

## 2 MANAGEMENT AND ORGANIZATION

The original SNS proposal had each of the participating laboratories responsible for the control system for its individual part of the machine – Berkeley for the Front End controls; Los Alamos for the linac controls; Brookhaven for the Ring controls, etc. While this model insured a tight coupling between individual control systems and the subsystems they controlled by making the partner laboratories responsible for both, there was risk that the ultimate integration of disparate control systems would have been difficult or impossible; and there was nothing in the model that allowed for the development of the "global systems" which are common to all – the network, the timing and synchronization system, the equipment protection system, a common control room, etc.

Eventually, a management model evolved that included the "Integrated Control System" (ICS) at the same organizational and reporting level as each of the six principle facility components – Front End, Linac, Ring, Target, Instruments and Conventional Facilities (the Physical Plant.) Each partner laboratory has a controls team and controls team leader that reports to the central (ORNL) controls team management. The global systems are themselves distributed, but managed centrally from Oak Ridge. This arrangement facilitates standardization and eventual integration, but requires that more effort be made to assure that subsystem developers and partner laboratories pay attention to the requirements, schedule imperatives and integration of their parts of the control system, which have become someone else's responsibility and are in someone else's budget.

One strategy to encourage integration at this level has been to include the subsystem controls schedule as a part of each system "sub-project" schedule, managed and "statused" at the partner laboratories and fully integrated with their schedules. The goal is to make subsystem designers more conscious of the control system support they require, and when. This is always a problem, and it is not clear that we have done any better than is usual. The downside is that there exists no separate control system schedule maintained centrally by the controls team. Understanding and reporting of schedule status and cost performance is a complicated and manpower-intensive effort requiring the integration of six different reports.

An important key to success in any project is good communication. This is rendered even more difficult, and more important, in a collaborative project such as SNS. The controls team has attempted to mitigate this problem with regular (weekly) teleconferences involving the controls team leaders at each of the participating laboratories. Those meeting originally included the use of "NetMeeting," but that became problematic with increased computer security at the DOE laboratories. Indeed laboratory firewalls have made all exchange of technical information awkward and inconvenient at best, impossible at worst. SNS has installed state-of-the-art videoconferencing facilities at each of the partner laboratories. These are used extensively, and are invaluable; however nothing satisfactorily takes the place of face-to-face discussion, and travel is an inevitable but necessary (and expensive) concomitant to successful collaboration. The SNS controls team was able to take advantage of this conference to have 23 members present at a controls team meeting. What might be a weekly occurrence under more conventional circumstances may be the first and last occasion for this group to be together for the duration of the project.

## 3 STANDARDS

Controls team leadership in the area of standardization has been a model for the rest of the project. Implementation and enforcement of standards in several areas, including software, hardware, screen design and device and signal naming was recognized very early as the linchpin of our integration approach.

### 3.1 Software

Software standardization has been difficult. In order to insure uniformity across all developed software, the SNS project negotiated project-wide licensing agreements. Some local sales organizations, however, have been reluctant to recognize these contracts, feeling that they have not got their "piece of the pie." This is of course a corporate problem, but one that nonetheless has resulted in delays and frustration for SNS implementers.

The most obvious, most important and most successful standard is the uniform use of EPICS for all subsystem controls. Contrary to tradition even in EPICS laboratories, this includes both the conventional facilities and the target control systems, where integration, often late, loose and ad hoc, was deemed important from the outset. Training was required for both commercial firms and partner laboratories not familiar with EPICS. Current development is taking place under EPICS v3.13, however the target version will be v3.15, which will have a number of capabilities added specifically for SNS [9].

EPICS itself allows a number of choices, and a suite of standard EPICS tools has been selected. This toolkit includes the "Extensible Display Manager" (EDM), developed for EPICS at the Oak Ridge Holifield facility, and which is being further developed in collaboration with the SNS controls team. EDM was chosen for easier maintenance and extensibility than competing EPICS display managers, and tools have been developed to translate screens developed in two of these: "MEDM" and "DM2K."

Working with the operations team, SNS has standardized on layouts and color use for operator screens. The EDM color rules capability facilitates this, allowing predefined colors to be selected by names such as: "linac background."

Linux has been chosen as the operating system for development, as well as console and high-level server applications, and nearly all utilities are available in this environment. Unfortunately VxWorks, the standard EPICS IOC kernel, still must be developed under Solaris, so the standard is not universal.

EPICS is oriented to individual signals, and does not provide a higher-level "device" view of the accelerator convenient to accelerator physics programmers. SNS is using a class library known as XAL [10] to address this requirement. In addition, temporary "ad hoc" programs can be written using Java-based Python scripts or in Matlab, either of which have direct access to any process variable in the control system.

The project early agreed on the use of Oracle™, with the goal of a fully integrated technical database relating device and signal tables for the generation of the EPICS distributed databases, lattice and modeling data for physics use, technical data on all equipment for tracking and maintenance, magnet measurement and other calibration data, a cable database, and more. A success for standardization? Well, not really – at least not yet. The usual difficulties arose. Many specialized databases were already in use. The schema did not correctly address all the issues in all of the diverse areas. Engineers were unwilling to give up control of their already useful databases to a central, and not always responsive, authority. Tools could not be centrally developed at the same rate as tools being developed in parallel at the partner laboratories. The result has been a struggle to incorporate existing databases and, so far, only moderate success in the use of what has become known as the Grand Unified Relational Database (GURD).

The most important tool for standardization and eventual integration of software developed at the partner laboratories is the requirement that all software be maintained and configuration managed out of a CVS repository located at Oak Ridge. The initiative has been moderately successful, and in spite of some resistance to developing in an environment far from home, more and more distributed developments are being deposited in the central repository. This should assure that all otherwise independent developments are using the same versions of the same tools, and avoid a potential integration nightmare when they all come together for final commissioning.

## 3.2 Hardware

SNS has facilitated the use of hardware standards by means of "Basic Ordering Agreements (BOAs)," which allow all partners, subcontractors and vendors to purchase selected standards at project-negotiated prices. This has worked fairly well, although intervention by the project is frequently required when agents for selected vendors do not or will not recognize negotiated prices as applying to them.

*PLCs:* The SNS control system makes far greater use of commercial Programmable Logic Controllers (PLCs) than is traditional in EPICS-based systems. PLCs are used for subsystems that must be kept operating whether or not the rest of the control system is needed, such as the cryogenic and vacuum control systems. In addition, the inclusion of traditionally PLC-based process systems such as those for conventional facilities and the target added even more PLC-based systems. SNS selected the Allan-Bradley ControlLogix™ family of PLCs for these applications.

Far more difficult has been the imposition of standards for the programming of PLCs. This is in part because many PLCs are in fact vendor-provided, often with pre-existent software. Related to programming, is the problem of divergent approaches to the use of PLCs. The guideline has been to use PLCs for interlocks only – whatever is needed to keep systems running safely, even when the EPICS control system might be down. All other functionality, such as automated procedures and operator displays, were to be implemented in the IOCs. This practice has not been universally adhered to.

*IOCs.* SNS has standardized on the Motorola 2100 Power PC series of processors for its distributed IOCs. From this family, an appropriate configuration can be selected for the application. An adapter card allows the same processor to be used for both VME and VXI applications. BOAs have also been established for VME and VXI crates: Dawn for 7 slot VME crates; Wiener for 21 slot crates and Racal for VXI.

*Racks.* A BOA has also been put in place for standard, 19" equipment racks. These may be configured as required with doors, side-panels and/or other accessories. The concept of a "rack factory" for assembly of equipment racks is not new – in house rack assembly facilities were established at SLAC for PEP II and at Jefferson Lab for CEBAF. SNS has signed a "rack-factory" contract with an electronic assembly company close to the SNS site to allow equipment designers, only if they choose, to have their equipment racks assembled at this facility on a "task-order" basis.

Racks would seem to be simple indeed. However even here imperfect communication exacerbated by distance resulted in a serious misunderstanding of significant consequence. Draftsmen at one partner laboratory misinterpreted the depth of the specified standard rack. As a result, an entire facility was laid out with racks that were too small. When this was eventually caught, all rack space had been allocated, some equipment would not fit the smaller rack and there was no room to add new rows of racks. The resulting compromise was ugly and not entirely satisfactory. This problem would likely not have occurred if all the principals were in one place.

## 3.3 Names and Database

Perhaps the first standard agreed by the partner laboratories was for signal and device naming. It has also given the most trouble. An apparently simple hierarchical standard was defined as shown below:

**SystemName:DeviceName:SignalName**

The names were to be mnemonic, long if needed for clarity, and optimized for operations. Instantiation schemes were defined for the linac and ring. Example lists of lattice devices were given. A document was signed and approved by all – so early in the project that no one needed it or used it. A state of euphoric innocence prevailed. Then reality intervened.

The concept hierarchically related each "signal" to a corresponding "device." The originators of this hierarchical idea intended that that device be an "accelerator" concept, such as a quadrupole, or a cavity, or a klystron; and that the signal be one of its observable properties. However signals can reasonably be associated with other types of devices – the transducer that produces it for example, or the VME ADC module that it goes to. All of these devices also need a place in the database for purposes of tracking and maintenance. The naming standard originators intended that these devices would be related to the signal using the relational database (although the original schema did not in fact do this), but that the signal *name* would use the "accelerator" device. As the design proceeded, poor communication, exacerbated by the difficulty of quickly detecting misunderstandings across several national laboratories, resulted in names being created that used any and all possible related devices in the signal name.

Different teams in different laboratories tried to apply the standard to their own subsystems. The scheme, which worked well and easily for lattice devices, was not so easy to apply to off-lattice devices such as a cryoplant or a cooling system. The hierarchical approach was foreign to engineers trained in the process control industry, who wished to relate SNS names to PLC "tag-names" formulated according to industrial standards. Engineers disagreed over what belonged in the device and signal fields, and how to apply the instantiation rules. Finally, a desire to contain all SNS technical information in an Oracle-based relational database, and to use this database (among other things) to produce the distributed control system database, imposed new requirements of parsability on the names that had not been considered in the standard.

By now, "official" names using several different interpretations of the original standards document appeared on drawings, screens and in documents and prototypical databases. This situation still prevails. Some names have been changed to conform to the intent of the standard, but this has not been done where an adverse schedule impact would have resulted. It may become necessary to make some changes later in the project, which will be both expensive and painful.

## 4 INDUSTRIAL PARTICIPATION

A controversial goal of the SNS approach was to fully integrate the "Conventional Facilities" controls from the outset. It is often the experience that these physical plant control systems (HVAC, power, etc) are provided by the general building contractor using technology quite different and incompatible with the accelerator control system. Later, during operations, it is found that process variables from these systems are needed in the control room, for observation or for correlation, and they are not easily accessible. Not without considerable controversy and opposition, SNS mandated that the conventional facility controls should be implemented from the outset in EPICS. To conform to traditional practice, this EPICS-based system would use PLCs at the I/O layer, and be implemented by a commercial contractor familiar with industrial control systems.

The Sverdrup Technologies controls team based in Tullahoma, Tennessee was awarded this contract. A weeklong EPICS training session was set up at Tullahoma, and this team is currently developing the distributed databases and human interface screens using the SNS-standard EPICS tools. As EPICS is based upon the same ideas as industrial control systems (the "I" in EPICS stands for "Industrial") the Sverdrup team seems quite at home in this environment, and is progressing well. They are familiar with PLC technology as well, and so can produce a fully integrated system, top to bottom. It is the plan to use the same approach and team to deliver both the EPICS and PLC portions of the Target control system. In addition to the obvious advantage of seamless integration of conventional with accelerator controls, this approach has made available an experienced commercial EPICS-trained team, which will be available later in the project to assist when there are resource shortages or schedule "crunches."

For more on the details of this arrangement, both technical and contractual, see reference [3].

## 5 HANDOFF

A particularly interesting challenge for the collaboration is the development and eventual implementation of plans to hand over to the SNS engineers and physicists at Oak Ridge complex subsystems developed at the partner laboratories. This applies to all subsystems, and is especially challenging for controls, where the systems include both hardware and software that might never have been fully integrated where they were designed. The project has developed a "Lead, Mentor, Consult" model for the handoff process, in which the partner laboratory responsible for the design of a subsystem takes a lead role for the design and for the installation and testing of the first subsystem; then allows Oak Ridge personnel to install and test the next subsystems, while taking an active mentoring role; and finally returns home to leave installation and testing of later subsystems to Oak Ridge personnel, while remaining available for consultation if needed. A detailed installation plan is in place that adopts this approach.

Two "facts of life" have made it difficult to implement this plan in an entirely rational and consistent manner. First, the SNS budget plan did not adequately account for anticipated pre-operations expenses. This resulted in pressure to move some money from the partner laboratories to SNS, thereby compromising their "lead" and "mentor" functions in some cases. Secondly, each of the partner laboratories have interpreted their responsibilities under this plan somewhat differently. The result is that while the controls team at Oak Ridge still expects considerable help with installation and testing of Linac subsystems, it expects to be more on its own in installing the Ring. These variations in approach have made it difficult to plan for staffing levels during the installation phase. One very great advantage of the collaborative model for project building is that the partner laboratories can serve as both source and sink for the extra staffing requirements of the construction phase. This benefit is being somewhat reduced by the need to increase staffing at Oak Ridge for pre-operations, at the expense of partner laboratory staff.

## 6 CONCLUSION

Construction at the Spallation Neutron Source site in Oak Ridge is proceeding on schedule. The "Front End" building will be complete at the end of May, and the Ion Source, RFQ and Medium Energy Beam Transport systems will be delivered in June and July of 2002. Installation of Linac components will begin in the fall of the same year. The project appears to be on track to deliver its first neutrons in December of 2005.

The control system should be ready to support installation, testing and commissioning of the various subsystems as they are delivered. The Front End already operates with beam at Berkeley, using a prototypical EPICS control system. "Hot Model" tests were supported by the controls team at Los Alamos, and prototypical controls subsystems are under development at the other partner laboratories. The refrigeration plant will be installed in the summer of 2002, and an EPICS control system will be ready. Control for "conventional facilities" is being developed under contract by a commercial vendor using EPICS, and each building, when handed over, will include an easily integrated EPICS control system for heating, ventilation, air conditioning, power, etc.

Like the facility itself, the control system for the SNS is being developed by a multi-laboratory collaboration. This presents unique management and organizational challenges. Attempts have been made to address these challenges in various ways, and with varying degrees of success. We are winning some, and losing some, and learning as we go; but the most important thing that we are learning for future projects is that there is no fundamental reason one cannot build and integrate a complex control system using many widely distributed partners. And have fun doing it.
.

## REFERENCES

[1] SNS Website: http://www.sns.gov/

[2] EPICS Website: http://www.aps.anl.gov/epics/

[3] This Conference: TUAP055. Use of EPICS for High-Level Control of SNS Conventional Facilities. J.K.Munro, Jr., J.C.Cleaves, E.L.Williams, Jr., D.J.Nypaver (ORNL), K.-U.Kasemir (LANL) and R.D.Meyer (Sverdrup Technologies)

[4] This Conference: WEAPO13. The SNS Cryogenic Control System: Experiences in Collaboration. W.H.Strong, P.A.Gurd, J.D.Creel and B.S.Bevans,

[5] This Conference: FRAT001. The SNS Timing System. B. Oerter, J.R.Nelson, T.Shea, C.Sibley

[6] SNS Preliminary Safety Assessment Document. Internal SNS Document.

[7] This Conference: TUAPO55. Plans for the Spallation Neutron Source Integrated Control System Network. W.R.DeVan and E.L.Williams

[8] PAC 2001 Proceedings, Session ROAB. SNS Beam Instrumentation and Challenges. T. Shea

[9] This Conference: THAPO14. Next Generation EPICS Interface to Abstract Data. J.Hill and R. Lange

[10] This Conference: THAPO60. SNS Application Programming Plan. C.M.Chu, J.Galambos, J.Wei and N.Malitsky